# Robust Cubature Kalman Filter for Dynamic State Estimation of Synchronous Machines under Unknown Measurement Noise Statistics

Yang Li[1], Senior Member, IEEE, Jing Li[1], Junjian Qi[2], Senior Member, IEEE, LiangChen[3]

[1]School of Electrical Engineering, Northeast Electric Power University, Jilin 132012, China
[2]Department of Electrical and Computer Engineering, University of Central Florida, Orlando, FL 32816, USA
[3]State Grid Heibei Economic Research Institute, Shijiazhuang 050041, China

Corresponding author: Yang Li (e-mail: liyang@neepu.edu.cn).

This work was supported in part by the China Scholarship Council (CSC) under Grant 201608220144, the National Natural Science Foundation of China under Grant No. 51677023.

**ABSTRACT** Kalman-type filtering techniques including cubature Kalman filter (CKF) does not work well in non-Gaussian environments, especially in the presence of outliers. To solve this problem, Huber's M-estimation based robust CKF (RCKF) is proposed for synchronous machines by combining the Huber's M-estimation theory with the classical CKF,which is capable of coping with the deterioration in performance and discretization of tracking curves when measurement noise statistics deviatefrom the prior noise statistics. The proposed RCKF algorithm has good adaptability to unknown measurement noise statistics characteristics including non-Gaussian measurement noise and outliers. The simulation results on the WSCC 3-machine 9-bus system and New England 16-machine 68-bus system verify the effectiveness of the proposed method and its advantage over the classical CKF.

## I. INTRODUCTION

### A. MOTIVATION

Accurate and reliable dynamic state information of synchronous machines plays a crucial role in real-time monitoring, protection, and control of power systems [1,2]. In addition, the emerging application of situational awarenessputs forward higher requirements for the status information acquisition as the system states evolve more complexly and quickly due to increasing cyber attacks [3] and high penetrations of renewable generations [4, 5]. However, some important synchronous machine state variables cannot be directly obtained. The successful industrial application of wide-area measurement system has recently made possible the estimation of all the state variables of a synchronous machine through the use of dynamic state estimations (DSE) [6]. Meanwhile, there is a higher-levelrequirement on DSE to ensure the safe and economic operation of a modern powersystem since its operation is increasingly close to the limits due to growing electricity demands and limited investments. Therefore, it is a pressing and challenging task to develop an effective DSE approach for synchronous generators.

### B. LITERATURE REVIEW

The term "dynamic state estimation" was first used in the 1970s [7], in which a Kalman filter technique was utilized to improve the performance of conventional quasi-static state estimation for power systems. In recent years, the studies on state estimators began to focus on a synchronous generator and its electromechanical transient model [8-10].In essence,this is a typical nonlinear filter problem. Up to now, there has been a significantly amount of studies on DSE of synchronous machines by using particle filters (PF) [11, 12] and variousKalman-type filtering algorithms, such asextended Kalman filter (EKF) [13-17], unscented Kalman filter (UKF) [18-24], and Cubature Kalman Filter (CKF) [3, 25, 26].The EKF is a classical nonlinear Kalman filter; the unscented transform-basedUKFprovidesreasonable filtering performance, but its convergence is dependent on the sampling methods of Sigma points [18, 19]; the CKF based on the spherical-radial cubature rule is an emerging nonlinear filter, which can give a systematic solution for high-dimensional nonlinear filtering issues. Extensive comparisons of all these Kalman-type estimators have been made from different perspectives, such as convergence, numerical stability, and computational complexity in [3, 16].

### C. LIMITATIONS AND CONTRIBUTIONS

The Kalman-type filtering techniques perform well under the Gaussian assumption [3,5,27]. However, the distribution of the measurement noise may not obey this assumption in practical applications. Recent research findings in [28] demonstrate that the errors in PMU measurements such as voltage phasors tend to follow non-Gaussian distributions with long tails such as Laplace distribution and often contain high-intensity noise realizations, called outliers, which could deteriorate the performance of the Kalman-type filtering approaches. Furthermore, the received measurements may be biased because of multiple reasons such as false data injection (FDI) attacks [3, 29]. Therefore, there is sustainable motivation for developing a robust filter that can work well in non-Gaussian environments and in the presence of outliers. In order to achieve such a goal, the work in [30] proposes a Generalized-Maximum Likelihood (GM)-UKF, in which a batch-mode regressing form is obtained via the statistical linearization to enhance the data redundancy, and thereby the form enables the GM-estimator to identify bad data and filter out unknown noises. However, when using the UKF, it is an essential but challenging task to generate Sigma points by using a scaled symmetric sampling strategy in case the state vector dimension is greater than 3, since there are three mutually influential parameters needed to be tuned in this step, while there is currently no consensus about the corresponding parameter selection principles.

In this work, a robust CKF (RCKF) based distributed DSE approach is developed to estimate the machine dynamic states by integrating the Huber's M-estimation theory with the CKF. Different from the GM-type estimator in [30], the proposed RCKF uses the robust M-estimation to detect outliers in measurements and then eliminates them by revising measurement noise variance matrix.

The main contributions of this work are as follows.

(1) A novel DSE algorithm, called RCKF, is proposed for synchronous machines by combining the Huber's M-estimation theory with the classical CKF, which has the ability to cope with the deterioration in performance and discretization of tracking curves when measurement noise statistics deviate from the prior noise statistics.

(2) The simulation results on the WSCC 3-machine 9-bus system and New England 16-machine 68-bus system demonstrate that the proposed approach is capable of addressing the DSE of synchronous machines under unknown measurement noise statistics.

(3) By sacrificing computational efficiency slightly, the proposed algorithm outperforms the conventional CKF under all the used noise conditions, including non-Gaussian measurement noise and outliers.

### D. PAPER ORGANIZATION

The remainder of this paper is structured as follows. The used estimation models are introduced in Section II. Section III presents the proposed RCKF in detail. Section IV gives case studies to examine the proposed approach. And finally, the conclusions are drawn in Section V.

## II. ESTIMATION MODELS

### A. SYNCHRONOUS MACHINE MODEL

Fourth-order transient model is a well-known generator model that has been extensively studied in previous literature [1, 9, 10, 21]. Its mathematical model is

$$\begin{cases} \dot{\delta} = \omega_0 \Delta\omega \\ \Delta\dot{\omega} = \frac{1}{T_J}\left[T_m - T_e - D\Delta\omega\right] \\ \dot{E_q'} = \frac{1}{T_{d0}'}\left[E_f - E_q' - (X_d - X_d')I_d\right] \\ \dot{E_d'} = \frac{1}{T_{q0}'}\left[-E_d' + (X_q - X_q')I_q\right] \end{cases} \quad (1)$$

$$\begin{cases} I_d = \frac{E_q' - U_t \cos(\delta - \phi)}{X_d'} \\ I_q = \frac{U_t \sin(\delta - \phi) - E_d'}{X_q'} \end{cases} \quad (2)$$

where $\delta$ and $\omega$ are respectively the rotor angle and speed; $E_q'$ and $E_d'$ are the q-axis and d-axis transient voltages; $T_m$ is the mechanical torque; $T_e$ is the electromagnetic torque; $D$ is damping coefficient; $T_{d0}'$ and $T_{q0}'$ are respectively the d- and q- axes transient time constants; $E_f$ is the field voltage; $X_q$ and $X_q'$ are q-axis synchronous and transient reactance; $X_d$ and $X_d'$ are the d-axis synchronous and transient reactance; $U_t$ and $\phi$ are the magnitude and phase angle of the generator terminal voltage, respectively; $I_d$ and $I_q$ are the d-axis and q-axis generator output currents.

The measurement equations are given by

$$\begin{cases} \omega^z = \omega \\ \delta^z = \delta \\ P_e^z = \frac{U_t^2}{2}\sin(2\delta - 2\phi)(\frac{1}{X_q'} - \frac{1}{X_d'}) + \\ \quad \frac{U_t^2 \sin(\delta - \phi)E_q'}{X_d'} - \frac{U_t^2 \cos(\delta - \phi)E_d'}{X_q'} \end{cases} \quad (3)$$

where $\delta^z$ and $\omega^z$ are the PMU measurements of rotor angle and rotor speed [31], and $P_e^z$ is the active power measurement.

The covariance of measurement noise $R_{k+1}$ is

$$R_{k+1} = diag\left(\sigma_{\delta_z}^2, \sigma_{\omega_z}^2, \sigma_{P_{e_z}}^2\right) \quad (4)$$

where $\sigma_{\delta_z}^2$ and $\sigma_{\omega_z}^2$ are the measurement variance of rotor angle and speed, and $\sigma_{P_{e_z}}^2$ is the variance of active power.

$$P_{e_z}^2 = \left(\frac{\partial P_e}{\partial U}\right)^2 \sigma_U^2 + \left(\frac{\partial P_e}{\partial \phi}\right)^2 \sigma_\phi^2 \quad (5)$$

where $\sigma_u = 0.2\%$, $\sigma_\phi = 0.2°$.

For ease of description, (1) is transformed the following continuous-time state space model [1, 16]:

$$\begin{cases} \dot{x} = F_C(x,u) + v_C \\ z = H_C(x,u) + w_C \end{cases}$$

$$x = \left[\delta, \omega, E_d', E_q'\right]^T \quad (6)$$

$$z = \left[\delta^z, \omega^z, P_e^z\right]^T$$

$$u = \left[T_m, E_f, U_t, \phi\right]^T$$

where $x$ is the state vector, $u$ is the input vector, and $z$ is the measurement vector; the subscript "C" denotes the continuous-time model; $F(\cdot)$ and $H(\cdot)$ are respectively the state transition and output functions; $v_C$ and $w_C$ are the process and output noise.

### B. PROBLEM FORMULATION

With a sampling interval $\Delta t$, a real-time DSE can be described as the following filtering problem: given inputs such as $T_m(j\Delta t)$, $E_f(j\Delta t)$, $U_t(j\Delta t)$, $\varphi(j\Delta t)$ for $j = 1, 2, \cdots$ we want to estimate the states of synchronous machines including $\delta(k\Delta t)$, $\Delta\omega(k\Delta t)$.

To perform state estimation via the discrete measurements, the continuous-time model in (6) is discretized into a discrete-time one as follows:

$$\begin{cases} x_{k+1} = F(x_k, u_k) + v_k \\ z_{k+1} = H(x_{k+1}, u_{k+1}) + w_{k+1} \end{cases} \quad (7)$$

where the subscript "k" is the moment at $k\Delta t$; $v_k$ and $w_{k+1}$ are the system process and measurement noises, and their covariance are denoted by $Q_k$ and $R_{k+1}$. In this study, the process noise $v_k$ is assumed to follow a Gaussian distribution, while the distribution of the measurement noise may not be Gaussian.

## III. ROBUST CUBATURE KALMAN FILTER

### A. CUBATURE KALMAN FILTER

The classical CKF was originally proposed in [32], which can be divided into two parts: time prediction and measurement update. In time prediction, CKF obtains a set of equally weighted state cubature points according to the spherical-radical rule. It can obtain predicted state variable and error variance matrix. In measurement update, the predicted state variable is updated by using measurements in order to improve estimation accuracy. The detailed process of the CKF is given as follows:

**1) Time Prediction**

Suppose that the estimation error covariance at time step $k$ is $P_{k|k}$ and its square-rooting matrix is $S_{k|k}$.

$$P_{k|k} = S_{k|k} S_{k|k}^T \quad (8)$$

$N$ cubature points of state variable are calculated.

$$X_{i,k|k} = S_{k|k} \zeta_i + \hat{x}_{k|k} \quad (9)$$

where $\hat{x}_{k|k}$ is the state estimatied valve at time step $k$. $X_{i,k|k}$ is the cubature point of $\hat{x}_{k|k}$, $\zeta_i = \sqrt{n}$, $i=1,2,\ldots 2n$, $n$ is the state vector dimension.

Through the state equations, the predicted values of the cubature points are obtained by

$$X_{i,k|k}^* = f\left(X_{i,k|k}, u_k\right) \quad (10)$$

where $X_{i,k+1|k}^*$ is the predicted value of $X_{i,k|k}$

The predicted values of state variable by weighted summation are obtained:

$$\hat{x}_{k+1|k} = \frac{1}{2n}\sum_{i=1}^{2n} X_{i,k+1|k}^* \quad (11)$$

where $\hat{x}_{k+1|k}$ is the predicted value of state variable.

The predicted error variance matrix of the state variable is

$$P_{k+1|k} = \frac{1}{2n}\sum_{i=1}^{2n} X_{i,k+1|k}^* X_{i,k+1|k}^{*T} - \hat{x}_{k+1|k}\hat{x}_{k+1|k}^T + Q_k \quad (12)$$

where $P_{k+1|k}$ is the predicted error variance matrix of state variable. $Q_k$ is the covariance of process noise.

**2) Measurement Update**

The square-rooting matrix of predicted error covariance is calculated according to

$$P_{k+1|k} = S_{k+1|k} S_{k+1|k}^T \quad (13)$$

where $S_{k+1|k}$ is its square-rooting matrix.

$N$ Cubature points of $\hat{x}_{k+1|k}$ are calculated by

$$X_{i,k+1|k} = S_{k+1|k} \zeta_i + \hat{x}_{k+1|k} \quad (14)$$

where $X_{i,k+1|k}$ is the cubature point of $\hat{x}_{k+1|k}$.

$N$ Cubature points of predicted measurement are

$$Z_{i,k+1|k} = h\left(X_{i,k+1|k}, u_k\right) \quad (15)$$

where $Z_{i,k+1|k}$ is the cubature point of predicted measurement.

The predicted measurement by weighted summation are

$$\hat{z}_{k+1|k} = \frac{1}{2n}\sum_{i=1}^{2n} Z_{i,k+1|k} \quad (16)$$

where $\hat{z}_{k+1|k}$ is the predicted measurement.

The innovation covariance matrix of measurement error is given by

$$P_{zz,k+1|k} = \frac{1}{2n}\sum_{i=1}^{2n} Z_{i,k+1|k} Z_{i,k+1|k}^T - \hat{z}_{k+1|k}\hat{z}_{k+1|k}^T + R_{k+1} \quad (17)$$

where $P_{zz,k+1|k}$ is the innovation covariance matrix. $R_{k+1}$ is the covariance of measurement noise.

The cross-covariance matrix $P_{xz,k+1|k}$ is obtained by

$$P_{xz,k+1|k} = \frac{1}{2n}\sum_{i=1}^{2n} X_{i,k+1|k} Z_{i,k+1|k}^T - \hat{x}_{k+1|k}\hat{z}_{k+1|k}^T \quad (18)$$

where $P_{xz,k+1|k}$ is the cross-covariance matrix.

The filter gain $W$ is obtained by

$$W_{k+1} = P_{xz,k+1|k} P_{zz,k+1|k}^{-1} \quad (19)$$

The estimated values of state variable are obtained by:

$$\hat{x}_{k+1|k+1} = \hat{x}_{k+1|k} + W_{k+1}(z_{k+1} - \hat{z}_{k+1|k}) \quad (20)$$

where $W_{k+1}$ is the filter gain. $\hat{x}_{k+1|k+1}$ is the state estimated value at time step $k+1$.

The estimation error covariance $P_{k+1|k+1}$ is updated for the next time step:

$$P_{k+1|k+1} = P_{k+1|k} - W_{k+1} P_{zz,k+1|k} W_{k+1}^T \quad (21)$$

**B. ROBUST CUBATURE KALMAN FILTER**

As a nonlinear filtering technique, the conventional CKF needs an accurate system model and noise statistical characteristics to work well. However, the measurement noise may not obey the Gaussian assumption in the actual scene. More importantly, the noise statistical characteristics might change due to the influence resulted from internal or external unknown factors during the estimation process. When outliers occur in PMU measurements, the covariance matrix of measurement noises $R$ will not consist with actual errors, and thereby the covariance matrix $P_{zz}$ in (17) is unavailable to reflect the deviation of predicted value, which eventually leads to an inaccurate estimation. These above factors, to a certain extent, limits the usefulness and practicality of the CKF in actual applications.

The robust M-estimation theory is an effective tool for addressing robust estimation against unknown noise statistics [33]. Through robust M-estimation, we can detect the outliers on state estimation and update the statistical characteristics of the measurement noise in real time, which makes CKF capable of adapting to the statistical characteristics of measurement noises. By combining the Huber's M-estimation theory with the classical CKF, the proposed RCKF can obtain the accurate DSE of synchronous generators with unknown measurement noise statistics.

The RCKF uses the Huber's M-estimation approach to obtain a revised covariance matrix of measurement noises $P_{zz}$ in (17). The corrected matrix $\bar{R}_{k+1}$ is substituted for the covariance of measurement noise in (17) as

$$\bar{R}_{k+1} = \bar{P}^{-1} \quad (22)$$

where the Huber's algorithm is utilized to calculate the equally weighted matrix $\bar{P}$.

Huber's M-estimation minimizes the cost function as

$$J(x_k) = \sum_{i=1}^{2n} \rho(r_i') \quad (23)$$

where $r_i'$ refers the $i$th component of the residual vector

$$r_i' = r_i / \sigma_{ri} \quad (24)$$

The $r_i$ and its standard deviation $\sigma_{ri}$ are calculated by

$$r_i = (z_{k+1} - \hat{z}_{k+1|k})_i \quad (25)$$

$$\sigma_{ri} = (P_{zz,k+1|k})_{i,i} \quad (26)$$

And the 'score function' $\rho(\cdot)$ is defined as

$$\rho(r_i') = \begin{cases} \frac{1}{2}r_i'^2, & |r_i'| \leq c \\ c|r_i'| - \frac{1}{2}c^2, & |r_i'| > c \end{cases} \quad (27)$$

where $c$ is a constant and is chosen as 1.5 in this paper.

Setting the partial derivative of (23) to be zero gives

$$\sum_{i=1}^{2n} \frac{\partial \rho(r_i')}{\partial (r_i')} \cdot \frac{\partial (r_i')}{x_{k,i}} \quad i = 1,2,\ldots n \quad (28)$$

where $x_{k,i}$ is the $i$th component of state vector. Denoting

$$w_i = \frac{\partial(r_i')}{r_i' \partial r_i'}, \quad i = 1,2,\ldots 2n \quad (29)$$

The formulation can be obtained

$$w_i = \begin{cases} 1, & |r_i'| \leq c \\ \dfrac{c}{|r_i'|}, & |r_i'| > c \end{cases} \quad (30)$$

Based on the above formulation, the equally weighted matrix $\bar{P}$ can be obtained as

$$\bar{p}\{i,i\} = \begin{cases} \dfrac{1}{\sigma_{i,i}}, & \left(\left|\dfrac{r_i}{\sigma_{ri}}\right| = |r_i'| \leq c\right) \\ \dfrac{c}{\sigma_{i,i}|r_i'|}, & (|r_i'| > c) \end{cases} \quad (31)$$

$$\bar{p}\{i,j\} = \begin{cases} \dfrac{1}{\sigma_{i,j}}, & (|r_i'| \leq c, |r_j'| \leq c) \\ \dfrac{c}{\sigma_{i,j}\max(|r_i'|,|r_j'|)}, & (|r_i'| > c, |r_j'| > c) \end{cases} \quad (32)$$

where $\bar{p}\{i,i\}$ and $\bar{p}\{i,j\}$ are diagonal and off-diagonal elements of matrix $\bar{P}$; $\sigma_{i,i}$ and $\sigma_{i,j}$ are diagonal and off-diagonal elements in measurement noise matrix $R_{k+1}$. Since the measurement noise covariance matrix is a diagonal matrix, $\sigma_{i,j}$ is zero. $r_i$ is the residual of measurement, $r_i^{'}$ is the standard residual error, $\sigma_{ri}$ is the mean variance of $r_i$.

## IV. CASE STUDIES

The proposed approach is tested on two systems, which are extracted from Power System Toolbox (PST) [34], and is compared with the original CKF in [25]. All the simulations are carried out on a PC with Intel Core i3-2330 2.20 GHz processor and 4 GB RAM.

The detailed simulation settings are listed as follow: 1) The simulation time step is set to 0.02s; 2) The standard deviations of the rotor speed (p.u.) and rotor angle(°) are set to 0.001 p.u. and 2°; 3) The standard deviations of the amplitude and phase angle of the output voltage are taken as 0.1% and 0.1°; 4) Each generator is equipped with a PMU at its terminal; 5) PMU measurements are assumed to be sampled at 50 samples/s.

### A. NOISE MODEL

We consider four types of noise: Gaussian white noise, Gaussian noise, Laplace noise, and Cauchy noise, which are respectively called noises 1-4.

*1) Gaussian white Noise and Gaussian Noise*

Gaussian noise is obtained by

$$f(x) = \frac{1}{\sigma\sqrt{2\pi}} e^{-\frac{(l-\mu)^2}{2\sigma^2}} \quad (33)$$

where $\mu$ and $\sigma$ are, respectively, the mean value and standard deviation of noise. When $\mu = 0$ and $\sigma^2$ is constant, we can obtain Gaussian white noise. If $\mu \neq 0$, we can obtain Gaussian noise.

*2) Laplace Noise*

Laplace noise with scale $s$ and mean $\mu$ is modeled as [3]

$$r_{Laplace} = \mu - s \times \text{sgn}(U_1)\ln(1-|U_1|) \quad (34)$$

where $s$ is the scale parameter ($s$ is taken as $\sigma/\sqrt{2}$) and $U_1$ is a random parameter that obeys uniform distribution in the sampling interval.

*3) Cauchy Noise*

Cauchy noise is generated by [3]

$$r_{Cauchy} = a + b\tan(\pi(U_2 - 0.5)) \quad (35)$$

where $a$ and $b$ are the location and scale parameters, and $U_2$ is a random parameter that follows uniform distribution in the sampling interval. Here, $a$ and $b$ are respectively chosen as $10\sigma$ and $\sigma$. The parameters are given in Table I.

TABLE I
NOISE TYPES AND PARAMETERS

| Noise type | Parameters | Standard deviation | Mean value |
|---|---|---|---|
| Noise 1 | degree(°) | 2 | 0 |
| | $\omega$ (%) | 0.1 | 0 |
| Noise 2 | degree (°) | 2 | 20 |
| | $\omega$ (%) | 0.1 | 1 |
| Noise 3 | degree (°) | 2 | 20 |
| | $\omega$ (%) | 0.1 | 1 |
| Noise 4 | degree (°) | 2 | 20 |
| | $\omega$ (%) | 0.1 | 1 |

Two widely-used classical indicators proposed in [35] are here utilized to evaluate the performance of the estimation.

$$\varepsilon_1 = \sqrt{\frac{\sum_{i=1}^{S_M}(\hat{x}_i - x_i^t)^2}{\sum_{i=1}^{S_M}(x_i^z - x_i^t)^2}} \quad (36)$$

$$\varepsilon_2 = \sqrt{\frac{1}{S_M}\sum_{i=1}^{S_M}\left(\frac{\hat{x}_i - x_i^t}{x_i^t}\right)^2} \quad (37)$$

where $\hat{x}_i$, $x_i^t$, $x_i^z$ are the estimation value, true value, and the measurement value of the sampling point $i$, and $S_M$ is the number of time steps.

### B. CASE 1: WSCC 3-MACHINE 9-BUS SYSTEM

A three-phase short-circuit permanent fault is applied at bus 5 to generate dynamic responses at $t$=1.2s. Then, the fault is cleared within the typical clearing time (5 cycles). The stimulation lasts for 20s. In real-world applications, bad PMU data must inevitably occur because of various causes such as impulsive noise, communication failures and potential/current transformer saturations, which leads to the severe deviation from the assumption that measurement noises obey the Gaussian distribution. To test the robustness of the proposed approach, two outlier scenarios are considered and tested, where a single outlier and a group of successive outliers are respectively added in the following two manners:

*Manner 1*: supposing that the rotor speed measurements are corrupted with 10% errors at the 6th second;

*Manner 2*: supposing that the rotor speed measurements are corrupted with 10% errors from $t$=2s to $t$=3s.

Taking generator 1 as an example, the estimation results under different types of noises are demonstrated in Figs. 1-8.

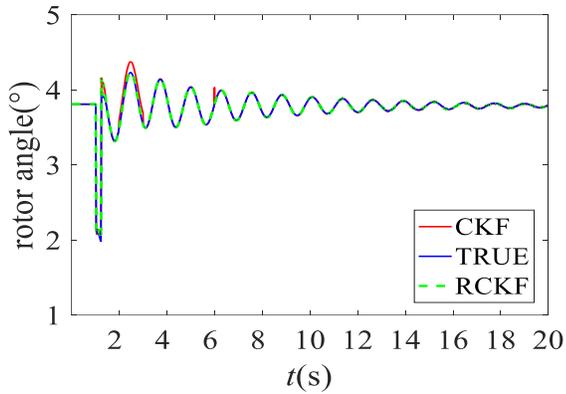

**Fig. 1   Rotor angle in Gaussian white noise**

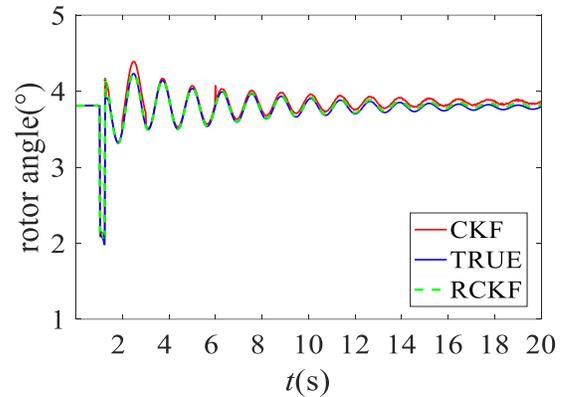

**Fig. 5   Rotor angle in Laplace noise**

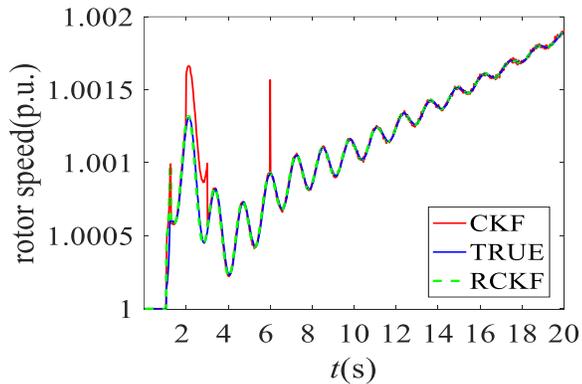

**Fig. 2   Rotor speed in Gaussian white noise**

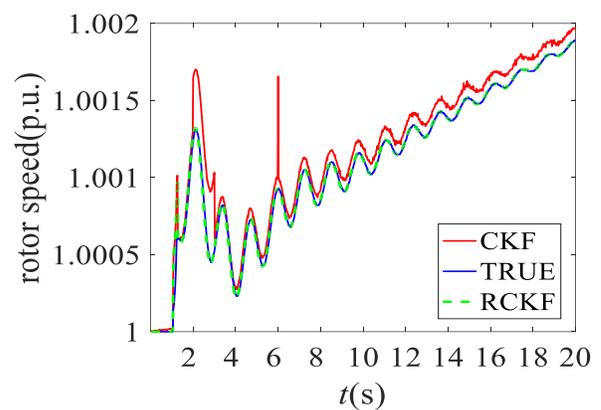

**Fig. 6   Rotor speed in Laplace noise**

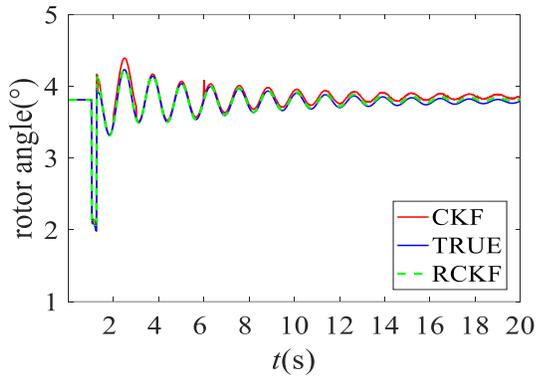

**Fig. 3   Rotor angle in Gaussian noise**

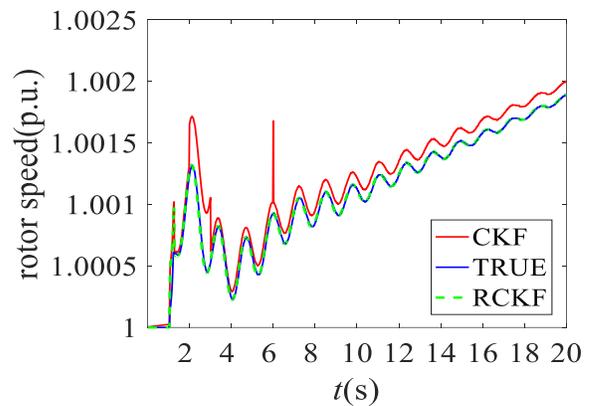

**Fig. 7   Rotor angle in Cauchy noise**

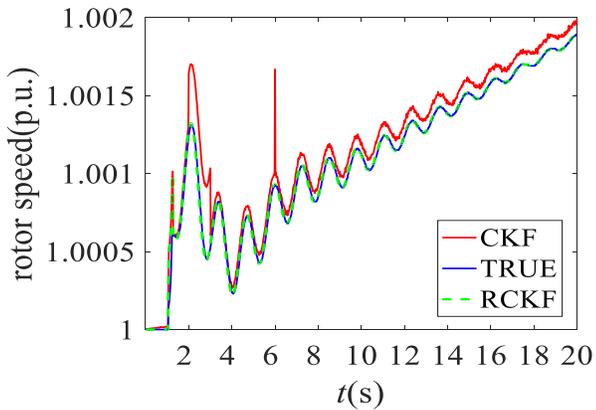

**Fig. 4   Rotor speed in Gaussian noise**

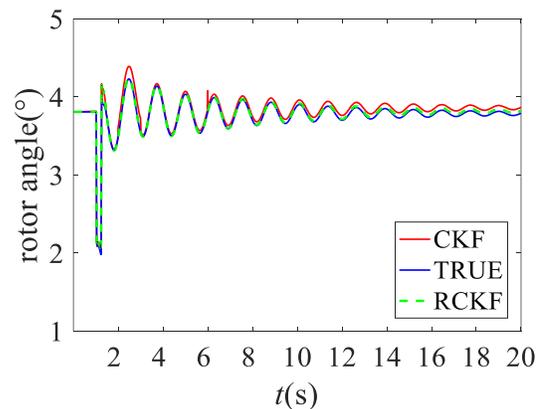

**Fig. 8   Rotor speed in Cauchy noise**

From Figs. 1-8, it can be seen that thefiltering performance the RCKF is superior to that of the CKF in the following two aspects. 1) Concerning the same noise: the RCKF has good robustnessin the presence of outliers; 2) Regarding various measurement noises, the RCKF shows good tracking and convergenceperformance; while the performance of the CKF declines markedly. This is because the RCKF can detect outliers and update the statistical characteristicsofthe measurement noisesviathe Huber's M-estimationtheory. By doing so, the estimated values of the RCKF can always converge to the true values accurately.

The quantitative comparisonresults of the two algorithms are demonstratedin Table II.

TABLE II
DYNAMIC STATE ESTIMATION INDEXES OF GENERATOR 2

| Noise type | index | Variables | CKF | RCKF |
|---|---|---|---|---|
| Noise 1 | $\varepsilon_1$ | $\delta$ | 0.0181 | 0.0085 |
| | | $\omega$ | 0.0041 | 0.0016 |
| | $\varepsilon_2$ | $\delta$ | 0.0248 | 0.0149 |
| | | $\omega$ | 0.0011 | 4.574e-04 |
| Noise 2 | $\varepsilon_1$ | $\delta$ | 0.0032 | 0.0011 |
| | | $\omega$ | 0.0046 | 0.0014 |
| | $\varepsilon_2$ | $\delta$ | 0.0388 | 0.0164 |
| | | $\omega$ | 0.0029 | 4.577e-04 |
| Noise 3 | $\varepsilon_1$ | $\delta$ | 0.0032 | 0.0012 |
| | | $\omega$ | 0.0049 | 0.0015 |
| | $\varepsilon_2$ | $\delta$ | 0.0387 | 0.0164 |
| | | $\omega$ | 0.0031 | 4.579e-04 |
| Noise 4 | $\varepsilon_1$ | $\delta$ | 0.0029 | 0.0014 |
| | | $\omega$ | 0.0045 | 0.0016 |
| | $\varepsilon_2$ | $\delta$ | 0.0369 | 0.0170 |
| | | $\omega$ | 0.0025 | 4.379e-04 |

From the above table, it can be observed that our approach outperforms the original CKF in the following two aspects. (1) Regarding the indicator $\varepsilon_1$: in the case of noise 1, theindicator values of the RCKF arerespectively increased by 52.7% and 60.9% compared with those of the CKF for rotor angle and rotor speed;in term of noise 2, theyareincreased by 65.6% and 69.5%; with noise 3, they are increased by 62.5% and 69.3%;with noise 4, they are increased by 51.7% and 64.4%. (2) In terms of the indicator $\varepsilon_2$: it can be seen thatthe filtering performance of the CKF significantly decreases in the case ofnon-Gaussian white noises (noises 2-4);while that of the RCKF has remained almost unchanged for all noises. This suggests that the RCKF can maintain good tracking performancesunder different noises.

It's worth noting that, as can be seen from (36), the indicator $\varepsilon_1$ of generator rotor angle $\delta$in Noise 1 is significantly greater than those in other noises since the variance of measurement errors is much less than other types of noises, as shown in (36). Therefore, the RCKF has good adaptability to unknown measurement noise statistics, and it can detect and eliminate the outliers in the measurements.

## C. CASE 2: NEW ENGLAND 16-MACHINE 68-BUS SYSTEM

This system includes 16 synchronous generators and 68 buses [3, 23]. A three-phase short-circuit fault is applied at bus 6 to generate a dynamic response at $t$=1s. The fault will be cleared at near and remote ends after 0.05s and 0.1s.The simulation lasts for 10s. One single outlier is superimposed on the 6th second, and 10 continuous outliers are superimposed from the third second.As the same in Case 1, two outlier scenarios are considered and tested, where a single outlier and a group of successive outliers are respectively added in the following two manners:

*Manner 1*: supposing that the rotor speed measurements are corrupted with 10% errors at the 6th second;

*Manner 2*: supposing that the rotor speed measurementsare corrupted with 10% errors from $t$=2s to $t$=3s.

Taking generator 1 as an instance, the dynamic state estimation results on the New England 68-bus system are demonstrated as Figs. 9-16 and Table III.

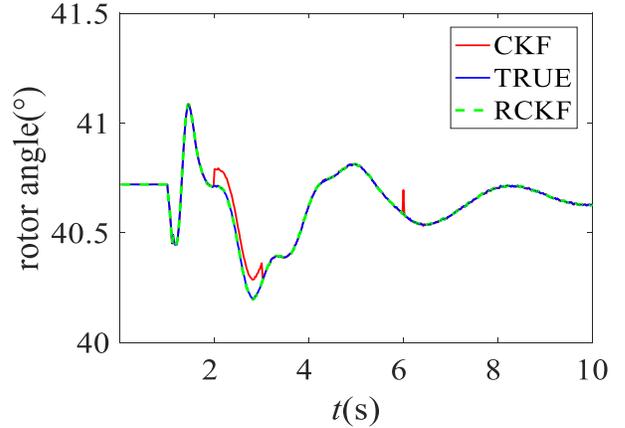

Fig. 9  Rotor angle in Gaussian white noise

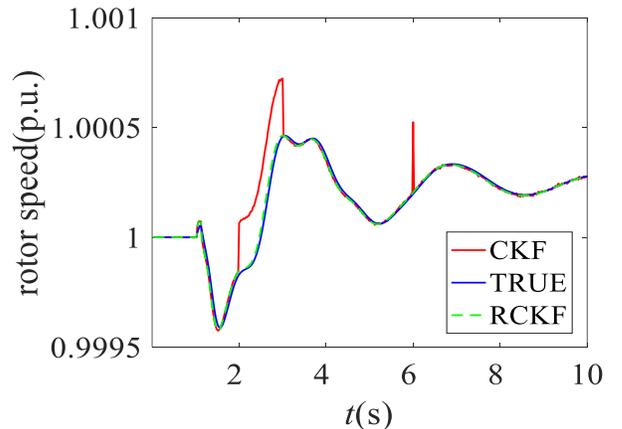

Fig. 10  Rotor speed in Gaussian white noise

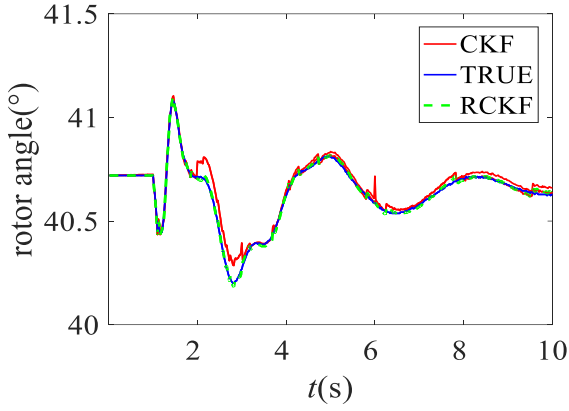

Fig. 11  Rotor angle in Gaussian noise

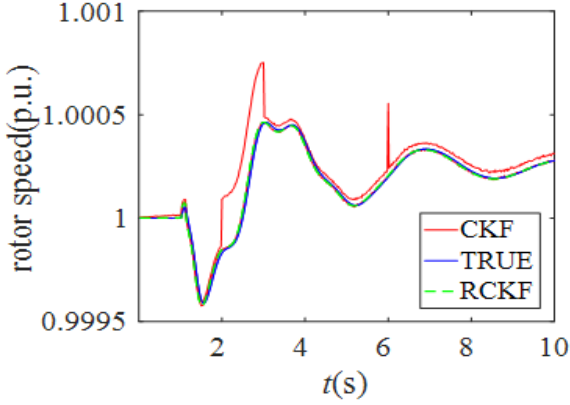

Fig. 12  Rotor speed in Gaussian noise

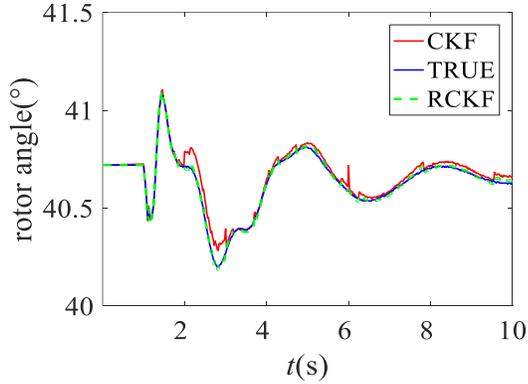

Fig. 13  Rotor angle in Laplace noise

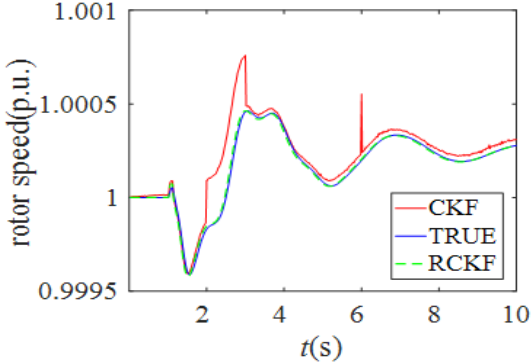

Fig. 14  Rotor speed in Laplace noise

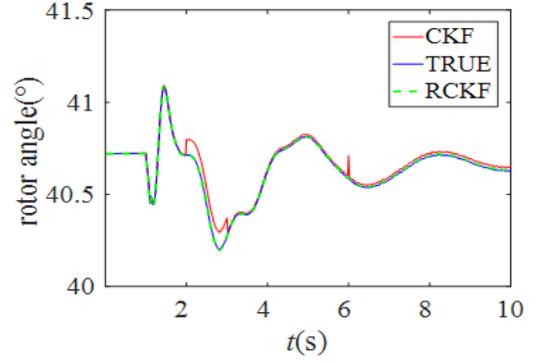

Fig. 15  Rotor angle in Cauchy noise

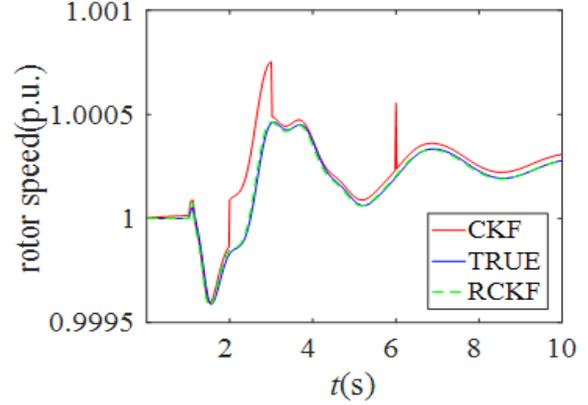

Fig. 16  Rotor speed in Cauchy noise

From Figs. 9-16, it can be observed that: the RCKF has better filtering performance than that of the CKF under various noises. Especially, in the case of non-Gaussian noises, the performance of the CKF becomes clearly poor while the RCKF can still maintain good estimation accuracy and convergence.

TABLE III
ESTIMATION RESULTS OF GENERATOR 1 IN 16 MACHINES SYSTEM

| Noises | index | Variables | CKF | RCKF |
|---|---|---|---|---|
| Noise 1 | $\varepsilon_1$ | $\delta$ | 0.014 | 0.0016 |
| | | $\omega$ | 0.0027 | 4.678e-4 |
| | $\varepsilon_2$ | $\delta$ | 0.0044 | 7.176e-4 |
| | | $\omega$ | 7.909e-4 | 2.124e-4 |
| Noise 2 | $\varepsilon_1$ | $\delta$ | 0.0017 | 1.800e-4 |
| | | $\omega$ | 0.0027 | 4.06e-4 |
| | $\varepsilon_2$ | $\delta$ | 0.0091 | 9.496e-4 |
| | | $\omega$ | 0.0012 | 2.121e-4 |
| Noise 3 | $\varepsilon_1$ | $\delta$ | 0.0018 | 1.801e-4 |
| | | $\omega$ | 0.0027 | 4.093e-4 |
| | $\varepsilon_2$ | $\delta$ | 0.0096 | 9.501e-4 |
| | | $\omega$ | 0.0012 | 2.3744-4 |
| Noise 4 | $\varepsilon_1$ | $\delta$ | 0.0020 | 2.257e-4 |
| | | $\omega$ | 0.0028 | 4.232e-4 |
| | $\varepsilon_2$ | $\delta$ | 0.0079 | 0.0064 |
| | | $\omega$ | 0.0011 | 2.126e-4 |

From TABLE III, it can be seen that: (1) Regarding indicator $\varepsilon_1$: the indicator value of the RCKF are respectively increased by 88.5% and 82.7% compared with those of the CKF for rotor angle and rotor speed under noise 1; they are increased by 89.4% and 84.9% under noise 2; they are increased by 90.0% and 73.7% under noise 3; they are increased by 88.7% and 83.8% under noise 4. (2) Concerning indicator $\varepsilon_2$: it can be observed that the performance of the CKF clearly deteriorates under non-Gaussian white noises; while that of the RCKF has remained almost unchanged over all noises. Based on these results, a conclusion can be drawn that the proposed approach also manages to perform the DSE of synchronous machines for a larger power system.

### D. COMPUTATIONAL EFFICIENCY

In order to properly evaluate the computational efficiencies of the proposed RCKF, the computation time for a single generator by using the CKF and RCKF in the above two cases are demonstrated in Table IV.

TABLE IV
CALCULATING TIMES OF THE CKF AND RCKF

| Cases | Noises | CKF (ms) | RCKF (ms) |
|---|---|---|---|
| Case 1 | Noise 1 | 0.301 | 0.362 |
| | Noise 2 | 0.323 | 0.335 |
| | Noise 3 | 0.325 | 0.341 |
| | Noise 4 | 0.312 | 0.335 |
| Case 2 | Noise 1 | 0.237 | 0.252 |
| | Noise 2 | 0.243 | 0.267 |
| | Noise 3 | 0.195 | 0.217 |
| | Noise 4 | 0.280 | 0.302 |

The results in Table IV show that the required computation times of the both algorithms are comparative, and the times are far less than a PMU sampling interval. This suggests that our approach is efficient enough to track the dynamic states of synchronous machines in real time, which is especially valuable for real-time applications such as emergency control. The computation time of the RCKF is only slightly more than that of the CKF. This is because additional computation is needed for the robust M-estimation to detect outliers and gross error in measurements and for eliminating them by revising the measurement noise covariance matrix.

### V. CONCLUSIONS

To resolve the lack of adaptability to unknown measurement noises using Kalman-type filtering techniques, a RCKF-based DSE approach for synchronous machines is proposed in this paper. By combing with the Huber's M-estimation theory and the original CKF, the proposed RCKF can detect outliers and gross errors, and thereby eliminate them by revising measurement noise variance matrix, which yields a more stable robust estimation. Finally, the simulation results of the WSCC 3-machine 9-bus system and New England 16-machine 68-bus system show that the presented approach has good robustness with outliers and good adaptability with unknown measure noises. More importantly, the filtering performances of the RCKF are far superior to those of the CKF against all types of noises used in this work.

Our future work will focus on extending this study to extensive potential applications in a whole power system. In addition, more realistic modeling techniques such as model uncertainties [36] and unknown inputs [37] will be incorporated to improve the practicality of our approach. Another interesting topic is to use the proposed algorithm for solving other estimation problems in engineering, such as state of charge estimation of battery storage [38, 39] and state estimation in combined heat and power networks [40].